\documentclass{article}
\usepackage{amssymb}

\usepackage{graphicx}
\usepackage{amsmath}

\begin{document}

\title{\textbf{Varying G, accelerating Universe, and other relevant
consequences of a Stochastic Self-Similar and Fractal Universe}}
\author{G.Iovane\footnote{iovane@diima.unisa.it}\\
Dipartimento di Ingegneria dell'Informazione e Matematica
Applicata,\\ Universit\'a di Salerno, Italy.}

\date{}

\maketitle

\begin{abstract}
In this paper the time dependence of G is presented. It is a
simple consequence of the Virial Theorem and of the
self-similarity and fractality of the Universe. The results
suggest a Universe based on  El Naschie's $\epsilon ^{(\infty )}$
Cantorian space-time. Moreover, we show the importance of the
Golden Mean in respect to the large scale structures. Thanks to
this study the mass distribution at large scales and the
correlation function are explained and are natural consequences of
the evaluated varying G. We demonstrate the agreement between the
present hypotheses of segregation with a size of astrophysical
structures, by using a comparison between quantum quantities and
astrophysical ones. It appears clear that the Universe has a
memory of its quantum origin. This appears in the G dependence
too. Moreover, we see that a $G=G(t)$ in El Naschie's $\epsilon
^{(\infty )}$ Cantorian space-time can imply an accelerated
Universe.
\end{abstract}

\newpage

\section{Introduction}
Observation shows a structure of a Universe with scaling rules,
where we can see globular clusters, single clusters or
superclusters of galaxies, in which stars can be treated as
massive point-like constituents of a universe mad of dust.

In a previous paper \cite{Iovane}, starting from an universal
scaling law, we showed its equality with the well--known Random
Walk equation or Brownian motion relation that was firstly used by
Eddington \cite{Elnaschie3}, \cite{Sidharth},\cite {capozziello}.
Consequently, we arrived at a self-similar Universe.
It was firstly considered by the Swedish Astronomers Charlier \cite{Rees}%
. Moreover, this law coincides with the Compton wavelength rule
when we just consider a single particle, for instance an electron.
By taking into account this generalization of Compton wavelength
rule, the model realizes a segregated Universe, where the sizes of
astrophysical structures can fit the observations (e.g. COBE,
IRAS, and surveys of large scale structures \cite{Lapparent}). The
idea, that a rule can exist among the fundamental constants, was
presented by Dirac and by Eddington--Weinberg, but these rules
were exact at Universe scale or subatomic scale. Here, a scale
invariant rule is presented. Thanks to this relation the Universe
appears self-similar and its self similarity is
governed by fundamental quantum quantities, like the Plank constant \textit{h%
}, and relativistic constants, like the speed of light \textit{c}.

It appears that the Universe has a memory of its quantum origin as
suggested by R.Penrose with respect to quasi-crystal
\cite{Penrose}. Particularly, it is related to Penrose tiling and
thus to $\varepsilon ^{(\infty )}$ theory
(Cantorian space-time theory) as proposed by M.S. El Naschie \cite{Elnaschie1}%
,\cite{Elnaschie2} as well as in A.Connes Noncommutative Geometry
\cite {Connes}.

In the present work, some ideas are presented about the
segregation of the Universe. In
particular, we analyze the scale invariant law $R(N)=\frac{h}{Mc}N^{\alpha }$%
, where $R$ is the radius of the astrophysical structures, $h$ is
the Planck constant, $M$ is the total Mass of the self-gravitating
system, $c$ the speed of light, $N$ the number of nucleons into
the structures and $\alpha \simeq 3/2$ is linked with the Golden
Mean. Consequently, thanks to the Virial Theorem we deduce the
gravitational constant $G$ and show that it is a function of the
time and the matter in the Universe (trough the number of
nucleons). Here our expressions agree with the Golden Mean and
with the gross law of Fibonacci and Lucas
\cite{Cook},\cite{Vajda}. We will see the effects of a non
constant G in stochastic self-similar Universe. In particular, we
analyze the cosmological density, the homogeneity of the Universe,
the implications coming from the Hubble's law, the correlation
function. We also deduce that a stocastic self-similar Universe or
equivalently an $\varepsilon^{(\infty)}$ Cantorian space-time
naturally imply an accelerated Universe. The paper is organized as
follows: we find the astrophysical scenario in Sec.2; Sec.3
presents a short review of definitions and properties for classic
and stochastic self-similar random processes; Sec.4 is devoted to
studying the effect of the Virial Theorem on the gravity
parameter-constant $G$; in Sec.5 we see how an accelerated
Universe comes from the Hubble's law in the context of stochastic
self-similar Universe; in Sec.6 we analyze some fundamental
consequences and finally conclusions are drawn in Sec.7.

\section{The Scenario}

As it well known luminous matter appears segregated at different
scale; in particular, we can distinguish among globular clusters,
galaxies, clusters
and superclusters of galaxies through their spatial dimensions \cite{Binney}%
, \cite{Vorontsov}.

It is interesting to note that if we write:
\begin{equation}
R(N)=\frac{h}{Mc}N^{\alpha },
\end{equation}
with $\alpha =3/2,$ for $M=M_{G}\thicksim 10^{10\div 12}M_{\odot }$ and $%
N=10^{68}$(this is approximately the number of nucleons in a
galaxy), we reproduce exactly $R\thicksim 1\div 10kpc$.

In general, we can evaluate the number of nucleons in a
self-gravitating system as
\begin{equation}
N=M/m_{n},
\end{equation}
where $N$ is the number of nucleons of mass $m_{n}$ into
self-gravitating
system of total mass $M$\footnote{%
In the present analysis the mass difference between proton and
neutron is not relevant such as it will be shown below. The mass
of nucleons is much larger than the mass of electrons,
$m_{p}=1836m_{e}$; therefore we can neglect the mass of
electrons.}. Then, we obtain the relevant results recalled in
Table 1. In the second column the number of evaluated nucleons is
shown, while we find the expected radius of self-gravitating
system in the last column.
\begin{eqnarray*}
&&\frame{$
\begin{array}{lll}
\text{\textbf{Sys Type}} & \text{\textbf{N.of Nucleons}} & \text{\textbf{%
Eval. Length}} \\
\text{Glob. Clusters} & N_{G}\thicksim 10^{63\div 64} &
R_{GC}\thicksim
1\div 10pc \\
\text{Galaxies} & N_{G}\thicksim 10^{68} & R_{G}\thicksim 1\div 10kpc \\
\text{Cluster of gal.} & N_{CG}\thicksim 10^{72} & R_{CG}\thicksim
1h^{-1}Mpc
\\
\text{Superc. of gal} & N_{SCG}\thicksim 10^{73} &
R_{SCG}\thicksim 10\div 100h^{-1}Mpc
\end{array}
$} \\
\text{Table 1} &:&\text{ Evaluated Length for different
self-gravitating systems}
\end{eqnarray*}
%\bigskip
By comparing the last column in Table 1 with the observed values,
we see a full agreement between the observed and theoretical
radius. It is obvious that if we have only one constituent (e.g.
$N=1$), like a proton or an electron, the relation (1) is the
standard and well--known Compton wavelength. Consequently, as
macroscopic system, our Universe shows a sort of quantum and
relativistic memory of its primordial phase. The choice to start
with $\alpha =3/2$ is suggested by statistical mechanics. By using
(2) the eq.(1) is strictly equivalent to
\begin{equation}
R(N)=l\sqrt{N},  \tag{1'}
\end{equation}
where $l=h/m_{n}c$. The relation (1') is the well--known Random
Walk equation or Brownian motion relation and it was firstly used
by Eddington \cite {Elnaschie3},
\cite{Sidharth},\cite{capozziello}.

In \cite{Iovane} we observed that $\alpha =3/2$  is a too rough
estimation if other interactions, in addition to gravity, are
relevant. For this reason, we considered stochastic self-similar
processes at atomic scale. These processes generalize the classic
ones. It was shown that the
nucleus scale is governed by a law like (1) but with a more complicated $%
l=l(N)$.

To determine the exact power law for astrophysical object, we
consider the mass and the radius of objects as known quantities
and evaluate the power law respect to the observed data. Let us
consider the relation
\begin{equation}
R(N)=\frac{h}{Mc}N^{x},
\end{equation}
where $x$ is the quantity to be determined.

Then, we obtain
\begin{equation}
x=\frac{\ln (RM/\alpha )}{\ln (N)},
\end{equation}
where $\alpha =h/c=2.2102209\times 10^{-42}Js^{2}m^{-1}$.

Table 2 summarizes the results in respect to the objects in the
length range $10pc<R<100h^{-1}Mpc$ and with a mass in the range
$10^{6\div 7}M_{\odot }<M<10^{17}h^{-1}M_{\odot }$.
\begin{eqnarray*}
&&\frame{$
\begin{array}{ll}
\text{\textbf{System Type}} & x \\
\text{Globular Clusters} & x_{GC}=1.5052\div 1.5084 \\
\text{Galaxies (Giant)} & x_{G}=1.4975\div 1.5273 \\
\text{Galaxies (Dwarf)} & x_{G}=1.5185\div 1.5435 \\
\text{Cluster of galaxies} & x_{CG}=1.5185 \\
\text{Supercluster of galaxies} & x_{SCG}=1.5180\div 1.5462
\end{array}
$} \\
&&\text{Table 2: Evaluated values of coefficient $x$ in power law} \\
&&\text{for astrophysical objects.}
\end{eqnarray*}
From Table 2, we note that, in the first approximation $x\simeq
1.5\simeq 3/2$. As suggested by El Naschie in various publications
\cite{Elnaschie5}, this is also in a close agreement with the
Fibonacci's numbers and the Golden Mean. In fact, we can write

\begin{equation}
R(N)=\frac{h}{Mc}N^{1+\phi }=\frac{h}{m_{n}c}N^{\phi },  \tag{3'}
\end{equation}

with $\phi =\frac{\sqrt{5}-1}{2}$$\ $.

If we make the hypothesis that relation (3) is a universal law,
then it has to be real at all scales. Table 3 summarizes the
results in respect to solar system objects.
\begin{eqnarray*}
&&\frame{$
\begin{array}{ccccc}
\text{\textbf{Object}} &
\text{\textbf{Radius(10}}^{6}\text{\textbf{m)}} &
\text{\textbf{Mass(Kg)}} & \text{\textbf{N}} & x \\
\text{Sun} & 6.96\times 10^{2} & M_{\odot } & 1.1892\times 10^{57}
& 1.4156
\\
\text{Mercury} & 2.439 & 3.2868\times 10^{23} & 1.9650\times
10^{50} & 1.4228
\\
\text{Venus} & 6.052 & 4.8704\times 10^{24} & 2.9112\times 10^{51}
& 1.4209
\\
\text{Earth} & 6.378 & 5.976\times 10^{24} & 3.5728\times 10^{51}
& 1.4206
\\
\text{Mars} & 3.3935 & 6.3943\times 10^{23} & 3.8229\times 10^{50}
& 1.4233
\\
\text{Jupiter} & 71.4 & 1.8997\times 10^{27} & 1.1358\times
10^{54} & 1.4205
\\
\text{Saturn} & 59.65 & 5.6870\times 10^{26} & 3.4000\times
10^{53} & 1.4232
\\
\text{Uranus} & 25.6 & 8.6652\times 10^{25} & 5.1806\times 10^{52}
& 1.4228
\\
\text{Neptune} & 24.75 & 1.0279\times 10^{26} & 6.1453\times
10^{52} & 1.4219
\\
\text{Pluto} & 1.1450 & 1.7928\times 10^{22} & 1.0718\times
10^{49} & 1.4270
\\
\text{Moon} & 1.738 & 7.3505\times 10^{22} & 4.3946\times 10^{49}
& 1.4254
\end{array}
$} \\
&&\text{Table 3: Calculated values of coefficient $x$ for solar
system objects.}
\end{eqnarray*}
By considering Table 3 we note the impressive constancy of
$x\thicksim 1.4$ for the planets of the solar system. The
discrepancy of $0.1$, in respect to the expected value $\alpha
=1.5$, could be an effect of the planets not being a
self--gravitating system. For the Sun this discrepancy is a little
bit worse than planets, probably due to not being a
self--gravitating system and because of the effects of nuclear
interactions in the interior of the Sun \footnote{In a recent
paper, Lynden-Bell and Dwyer have derived from first physical
principles a universal mass-radius relation for planets, white
dwarfs and neutron stars \cite{Lynden}. In the roughest
approximation, the proposed mass-radius relation for planets
reduces to $R\sim a_{0}\left(
M/m_{p}\right) ^{1/3}$ where $a_{0}=$\textit{%
%TCIMACRO{\UNICODE{0x127}}%
%BeginExpansion
h\hskip-.2em\llap{\protect\rule[1.1ex]{.325em}{.1ex}}\hskip.2em%
%EndExpansion
}$/m_{e}e^{2}$ which is equivalent to (1) when $\alpha =4/3.$
Then, at this scale, a stochastic self-similar process is more
appropriate than a self-similar one (see \cite{Iovane}). However,
also in this case the relation can be considered as a recasting
one of the Golden Mean or of the Fibonacci's law.}. A similar
approach was recently presented in \cite{Agop}; also in this work
the results reflect the Cantorian-fractal structure of the
space-time \footnote{A detailed and interesting study on the
quantization of the solar system was made by L.Nottale,
G.Schumacher and J.Gay \cite{Nottale5}.}.

\section{Classic and stochastic self-similar random process}

%\bigskip
Let $\Re $ be real space and $\gamma _{r}\in \Re ^{+}$, then we
define a self-similar (ss) random process for every $r>0,$
\begin{equation}
X(s)\overset{d}{=}\gamma _{r}X(rs),\qquad with\qquad s\in \Re ,
\end{equation}
where \ $\overset{d}{=}$\ denotes equality as distributions
\cite{Vervaat}.

The relation (5) is invariant under the group of positive affine
transformations,
\begin{equation}
X\rightarrow \gamma X,\qquad s\rightarrow rs,\qquad \gamma _{r}>0.
\end{equation}
Since\ $\gamma _{r}$\ satisfies the properties
\begin{gather}
\gamma _{r_{1}r_{2}}=\gamma _{r_{1}}\gamma _{r_{2}},\qquad \forall
r_{1},r_{2}>0, \\
\gamma _{1}=1,  \notag
\end{gather}
then it must have the form
\begin{equation}
\gamma _{r}=r^{-\delta },\qquad with\qquad \delta \in \Re .
\end{equation}
Thanks to (8) the relation (5) becomes
\begin{equation}
X(s)\overset{d}{=}r^{-\delta }X(rs),\qquad with\qquad s\in \Re .
\end{equation}
When a process satisfies (5) or (9), it is said to be self-similar or $%
\delta -$self-similar.

A generalization of self-similar random process is obtained by
replacing the deterministic scaling factor $\gamma _{r}=r^{-\delta
}$ in (5) or (9) with a random variable $\widetilde{\gamma
_{r}}\in \Re _{0}^{+}$. This variable is independent of the
process to which such a variable is multiplied. Then eq.(9)
becomes
\begin{equation}
X(s)\overset{d}{=}\widetilde{\gamma _{r}}X(rs),\qquad with\qquad
s\in \Re .
\end{equation}
D.Veneziano demonstrated in \cite{Veneziano} that
$\widetilde{\gamma_{r}}$
can also be written as $\gamma _{r}=r^{-\widetilde{\delta }}$ with $%
\widetilde{\delta }$ real random variable. Then, these kinds of
processes, called stochastic self-similar (sss) random processes
and the previous ones
(ss), can be treated in the same theory. Gupta and Waymire showed that for $%
0<r\leq 1$ the sss processes are dilations, while for $r>1$ the
sss processes are contractions \cite{Gupta},\cite{Gupta2}.

In \cite{Veneziano} the author proved the following relevant theorem: if $%
\widetilde{\delta_{r_{1}}}\overset{d}{=}\widetilde{\delta_{r_{2}}}$
for some $r_{1}\neq r_{2},$ then $\widetilde{\delta }$ must be a
deterministic constant $\delta .$ Then, one can treat ss and sss
random processes in a unique scheme. \newline Moreover, the author
gives many relevant properties and generalizations to a
$d$-dimensional space in the same paper, but we are not going to
consider these properties because they do not fit the objectives
of our paper (for more details see \cite{Veneziano}).

Presently it appears clear there is an agreement among (5), (10)
and (1), (1'). In fact, by defining the deterministic scaling
parameter $\gamma _{r}=r^{-1/2}$, we find
\begin{equation}
R(N)=\gamma _{r}R(rN);
\end{equation}
then our studies will explore a $1/2$-self-similar random
aggregation process.

By considering electromagnetic and nuclear interactions, relation
(10) becomes

\begin{equation}
R(N)=\widetilde{\gamma _{r}}R(rN),
\end{equation}
with $\widetilde{\gamma _{r}}$\ a random variable. In principle,
we have to expect a change from a deterministic scaling parameter
$(\gamma )$\ \ to a random one $(\widetilde{\gamma })$, due to
quantum treatment of nuclear and electromagnetic interactions.

\section{The Virial Theorem and the variation of G}

\noindent It is well known that the Virial Theorem can be
formalized as

\begin{equation}
2T+U=0,
\end{equation}

where $T$ is the Kinetic Energy and $U$ the Potential one. If we
consider a test particle we have $T=\frac{1}{2}mv^{2}$ and $%
U=-GMm/r$; consequently we obtain the well known expression to the
speed

\begin{equation}
v=\sqrt{\frac{GM}{r}}.
\end{equation}

If we consider $M=N$ $m_{n}$ and $v=\frac{dr}{dt}$ by integrating
on the range $\left[ 0,R\right] $ the following expression

\begin{equation*}
\int_{0}^{R}r^{1/2}dr=\int_{0}^{t}\sqrt{Gm_{n}N}dt
\end{equation*}

where $R$ is the radius of the structure with mass $M$, we easily
obtain

\begin{equation}
R=\overline{R}t^{2/3},
\end{equation}

with $\overline{R}=\left( \frac{3}{2}\right) ^{2/3}\left(
Gm_{n}N\right) ^{1/3}.$

Now if we impose that the previous expression is equal to the
eq.(1) we obtain

\begin{equation}
G=\overline{G}N^{1/2}t^{-2},
\end{equation}

where $\overline{G}=\left( \frac{2}{3}\right) ^{2}\frac{l^{3}}{m_{n}}$ with $%
l=h/m_{n}c$. In other words G is a function of the time and the
number of the constituents of the structures, i.e. $G=G(N,t)$. The
value  $\alpha =1/2$ coincides with $<d^{(0)}_{c}>$ in the El
Naschie theory of Cantorian fractal space-time. On the other hand,
we can also consider the value $d^{(0)}=\phi$ connected with the
Golden Mean and the Cantorian fractal structure of the space-time.
In this case we obtain the relation,
\begin{equation}
G=\overline{G}N^{3\phi -1}t^{-2}.
\end{equation}

\section{The Hubble's law and the accelerated Universe}
Starting from the Hubble's law

\begin{equation}
v(t)=H(t)R(t),
\end{equation}

we get
\begin{equation}
\frac{dv}{dt}=Hv+\frac{dH}{dt}R=\left( H^{2}+\frac{dH}{dt}\right)
R.
\end{equation}

Let us introduce the following parameter
\begin{equation}
q\equiv -\left( 1+\frac{1}{H^{2}}\frac{dH}{dt}\right) ,
\end{equation}

\bigskip named in literature decelerating parameter. Then we get

\begin{equation}
\frac{dv}{dt}=-qH^{2}R.
\end{equation}

By considering a spherical volume with a radius $R<R_{U}$ ($R_{U}$
is the radius of the Universe), large enough to obtain a regular
density, the gravitational force acting on an object posed on the
spherical surface becomes

\begin{equation}
\frac{dv}{dt}=-\frac{GM}{R^{2}},
\end{equation}

where $M$ is the mass which is inside the sphere. Consequently by
comparing the last two relation, we reach the result
\begin{equation}
q=\frac{GM}{H^{2}R^{3}}.
\end{equation}

It coincides with the well known expression $q_{0}=\frac{4\pi G}{%
3H_{0}^{2}}\rho _{0}$, when we consider $M=\frac{4}{3}\pi
R^{3}\rho _{0}$ (the index 0 means at the present).

We can evaluate the link between $H$ and $G$ as follows. By
substituting the expression (14) for $v$ coming from the Virial
Theorem in the Hubble's law (18), we easily get

\begin{equation}
H=\left( GM\right) ^{1/2}R^{3/2}.
\end{equation}

Consequently by taking into account the relations (2), (3) and
(16), we obtain

\begin{equation}
H=H(N,t)=\overline{H}N^{3/2}t^{-1},
\end{equation}

with $\overline{H}=\frac{2}{3}l^{3}$ with $l=h/m_{n}c$. If we use
(17) instead of (16), we obtain
\begin{equation}
H=H(N,t)=\overline{H}N^{3\phi }t^{-1}.
\end{equation}

\bigskip By evaluating $\frac{dH}{dt}$ and using the relations(20),
(25),and (30), we reach the result

\begin{equation}
q=-\left( 1-\frac{1}{\overline{H}N^{3/2}}\right) <0.
\end{equation}

If we use (26) instead of (25), we obtain the following relation,
which is linked with the Golden Mean
\begin{equation}
q=-\left( 1-\frac{1}{\overline{H}N^{3\phi}}\right) <0.
\end{equation}

The previous expressions suggest an accelerating Universe in
agreement with the observations. Recent measurements of a Type Ia
Supernova (SNe Ia), at redshift $z\sim1$, indicate that the
expansion of the present Universe is accelerated
\cite{Perlmutter},\cite{Garnavich},\cite{Riess},\cite{Bennett}. In
this sense, a stochastic self-similar and fractals Universe could
suggest the presence of dark energy pervasing the Universe. In its
simplest form, dark energy might well be Einstein's cosmological
constant in the form of a vacuum energy. On the other hand, recent
theories have been proposed including the possibility of slowly
evolving scalar fields (see quintessence models in
\cite{Ratra},\cite{Caldwell},\cite{Capozziello_q}). However the
future results of SNAP Collaboration (SuperNova Acceleration
Probe) coming from space-telescope give us the amount of dark
energy \cite{Snap}.

In conclusion when the observations convincingly demonstrated that
the expansion of the Universe is accelerated, a number of
theoretical investigations were started. Most explanations
suggested so far seem to belong to one of three categories: i)
assuming a nonzero cosmological constant
\cite{Carroll},\cite{PebRat}; ii) assuming a new scalar field
quintessence \cite{Ratra},\cite{Caldwell}; iii) or assuming new
gravitational physics. Here, we demonstrated that by using the
standard physics, that is the Virial Theorem and the Hubble's law
but on Cantorian space-time $\varepsilon^{(\infty)}$, the same
results can be achieved.

\section{Some consequences of the fundamental scale
invariant law}

We considered the Compton wavelength expression as a particular
case of a more general relation, which is true for all material
structures in the Universe. We discovered a fundamental relation
which demonstrates the self-similarity of the Universe. The
relations (1) or (3') show a Universe that has memory of its
quantum and relativistic nature at all scales. In this sense, the
Plank constant and the speed of light play a fundamental role in
giving a quantum and relativistic parameterization of the
structures. This reveals why the astrophysical structures and
organic matter have their particular lengths \cite{Elnaschie5}. In
what follows we analyze the consequences coming from the law (3)
or (3') and the relation (16) or (17).

\subsection{ the cosmological density $\Omega$}
The presented scale invariant law can be used to evaluate the
baryonic mass of the Universe. From (1) (with $\alpha =3/2$) we
have\footnote{%
\ As we have seen this is the best value when gravity is the only
relevant interaction.}
\begin{equation}
M_{U}=\left( \frac{Rc}{h}m_{p}^{3/2}\right) ^{2}=3.2841\times
10^{55}\text{ kg,}
\end{equation}
which corresponds to a number of nucleons of
\begin{equation}
N_{nucl}=\frac{M_{U}}{m_{p}}=1.9634\times 10^{82}.
\end{equation}
In the previous evaluation we considered a Universe with $%
R_{U}=6000Mpc=1.8516\times 10^{26}$ m. Let us introduce  the
critical density
\begin{equation}
\rho _{c}=\frac{3H^{2}}{8\pi G}=2\times
10^{-29}h^{2}\text{g/cm}^{3},
\end{equation}

evaluated with $H=100$ $h$ km s$^{-1}$Mpc$^{-1}$(where $0.5<h<1)$.

Clearly the approximation of a spherical Universe is too rough.
The self similarity of relation (3) or (3'), and the exponent
equal to 3/2 are the two fundamental ingredients of fractal
geometry. The scale invariant law lives in a fractal domain.

Therefore, the Universe has fractal dimension. Following
\cite{Mandelbrot}, \cite{Sylos}, \cite{Falconer} we can define the
fractal dimension as following
\begin{equation}
D=\underset{R\rightarrow \infty }{\lim }\frac{\ln (N(<r))}{\ln
(R)},
\end{equation}
where $N(<R)$ is the number of nucleons inside the radius $R$ and
$R$ is the radius of the structure. Thanks to (32), we can
estimate the fractal dimensions of all astrophysical structure and
of the Universe too. Table 4 summarizes these results.

\begin{eqnarray*}
&&\frame{$
\begin{array}{cc}
\text{\textbf{System Type}} & D \\
\text{Globular Clusters} & 3.61\div 3.66 \\
\text{Galaxies (Giant)} & 3.27\div 3.54 \\
\text{Galaxies (Dwarf)} & 3.18\div 3.39 \\
\text{Clusters of galaxies} & 3.20 \\
\text{Superclusters of galaxies} & 2.94\div 3.15 \\
\text{\textbf{Universe}} & 3.13
\end{array}
$} \\
&&\text{Table 4: Fractal Dimension of astrophysical objects}
\end{eqnarray*}
%\bigskip
From Table 4 it is very interesting to note a relative coincidence
of the fractal dimension of the Universe with the number $\pi$.

Taking into account the result
\begin{equation}
D=3.1329,
\end{equation}
it suggests a Universe whose spatial bound permeates the time
dimension.

If we also consider the time, then
\begin{equation}
D^{(4)}=4.1329.
\end{equation}
Theoretically speaking by assuming the limitation of measurement
accuracy the previous value can be the Hausdorff dimension, found
by El Naschie \cite {Elnaschie5}

\begin{equation}
D^{(4)}\cong \langle Dim~\varepsilon ^{(\infty )}\rangle
_{H}=4+\phi^{3} =4.236067977.  \tag{34'}
\end{equation}

On the other hand the value in (34) is also near to the continuous
$\Gamma$ distribution formula for $\varepsilon^{(\infty)}$; in
this case we obtain $D=2/ln(1/\phi)=2/ln(1.6180)=4.1562$.
\bigskip It is also interesting to note that $D^{(5)}=5.1329$, which is
connected with the fine structure constant, i.e

\begin{equation}
\left( D^{(5)}\right) ^{3}=\left( 5.1329\right) ^{3}\cong \alpha
_{0},
\end{equation}

as determined by El Naschie in \cite{Elnaschie5}.

Consequently, the density of the Universe is:
\begin{equation}
\rho _{U}^{fractal}=\frac{M_{U}}{4/3\pi
(R_{U})^{D}}=2.\,\allowbreak 134\,2\times 10^{-30}\text{g/cm}^{3},
\end{equation}
which is evaluated in the hypothesis of a spatial pseudo-sphere
Universe (see Fig.3). A similar result can be reach by using a
different approach based on the limit set of Klenian groups
\cite{Elnaschie4}. Therefore, $\rho _{U}<\rho _{c}$ indicates an
open universe, i.e. the gravitational interaction is not
sufficient to reverse the expansion of the Universe into a
contraction. This conclusion fully agrees with the present
observations \cite{Bernardis},\cite{Boomerang}. Moreover, the
cosmological density results in
\begin{equation}
\Omega =\frac{\rho _{U}}{\rho _{c}}=\frac{2.\,\allowbreak
134\,2\times 10^{-30}}{2\times 10^{-29}}=0.\,\allowbreak 11\ .
\end{equation}
We may mention at this point that El Naschie in \cite{Elnaschie5},
\cite{Elnaschie6} used the
dimensionless gravity constant $G$ to establish a Shanon-like entropy

\begin{equation}
S(G)= \frac{\ln G}{\ln 2}+1=\overline{\alpha }_{e\omega }\simeq
128,
\end{equation}

where $%
\overline{\alpha }_{e\omega }$\ is the coupling constant at the
Higgs-Electroweak in order to establish quantum gravity.\

The figures 1-2 show some examples of a pseudo-sphere Universe

%\begin{figure*}[tbp]
%\resizebox{8 cm}{!}{\includegraphics{fig5.EPS}} \hfill
%\parbox[b]{8 cm}{
%\caption{Some trivial examples of a spatial pseudo-sphere
%Universe.} \label{Figure1}}
%\end{figure*}
%
%\begin{figure*}[tbp]
%\resizebox{6 cm}{!}{\includegraphics{fig4.EPS}} \hfill
%\parbox[b]{6 cm}{
%\caption{A more realistic spatial pseudo-sphere Universe.}
%\label{Figure2}}
%\end{figure*}

\subsection{The homogeneous Universe}
The homogeneity of the Universe is also connected with the
self-similarity and with the parameter $\alpha =3/2$. If we
consider a constant number of galaxies per unit of volume, the
total number of galaxies $N$ along a direction and at distance $r$
grows like $r^{3}$. Consequently, by considering the galaxies with
an intrinsic luminosity $L$ the apparent luminosity $S$, that is
measured on the Earth, is

\begin{equation*}
S=\frac{L}{4\pi r^{2}},
\end{equation*}

then $r^{3}\varpropto S^{-3/2}$ and so

\begin{equation}
N(>S)\varpropto S^{-3/2}.
\end{equation}

This result is in agreement with the observation \cite{Peebles}.
The radio-sources and quasars appear in disagreement \ with the
previous result in (39) and suggest a greater value which tends to
1.6. For this reason we can suspect and propose

\begin{equation}
N(>S)\varpropto S^{-(1+\phi )},
\end{equation}

and consider (39) an approximation of the real value (40) linked
with the Golden Mean $\phi$.

\subsection{The correlation function and other connections with the Golden Mean}
It is well known that if we apply the Virial Theorem to a generic
distribution of galaxies with a correlation function $\xi (R)$, we
obtain the cosmic Virial Theorem, that is

\begin{equation}
<v^{2}>=\frac{1}{2}\frac{GM}{R},
\end{equation}

where $M=M(\xi (R))$. In particular, by considering a constant mass density $%
\rho $ on a spherical region, $M=\frac{4}{3}\pi \xi (R)\rho
R^{3}$. The correlation function has a power law with the exponent
$\gamma =1.8$,

\begin{equation}
\xi (R)=\left( \frac{\varsigma }{R}\right) ^{1.8},
\end{equation}

where for istance $\varsigma =5$ Mpc when $R<20$ Mpc \cite{Peebles}, and $%
\varsigma =26$ Mpc when $R<100$ Mpc \cite{Bahcall}. The relevant
exponent $\gamma $ can be derived from $\varepsilon ^{(\infty )}$
space-time. In fact, $\gamma
=10\phi ^{3}(1-\phi ^{3}),$ where $\phi $ is the Golden Mean. Moreover, $%
\gamma =10k$, where $k=\phi ^{3}(1-\phi ^{3})$ was invoked for
obtaining the dimensionless constant of gravity $\overline{\alpha
}_{G}=1.693(10^{38})$ by El Naschie \cite{Elnaschie4}. We mention
that starting from the expression

\begin{equation}
\overline{\alpha }_{G}=\frac{1}{G_{New}}\frac{\hbar c}{m_{Plank}},
\end{equation}

and by considering the (17) we obtain a time dependent $\overline{\alpha }%
_{G}$, that is

\begin{equation}
\overline{\alpha }_{G}\left( N,t\right) =\frac{1}{G_{0}N^{3\phi -1}}\frac{%
\hbar c}{m_{Plank}}t^{2}.
\end{equation}

Another interesting link between the Golden Mean and the
stochastic self-similar Universe is the following one. In the
context of the velocity of our galaxy with respect to the
Microwave Radiation Background, as showed by Peebles in \cite
{Peebles}, by applying the perturbation theory we can demonstrated
that the
velocity perturbation is proportional to the excess of the mass density $%
\rho $ in the Virgin cluster, that is

\begin{equation}
\frac{\delta v}{v}=\frac{1}{3}\left( \Omega \right)
^{0.6}\frac{\delta \rho }{\rho }.
\end{equation}

Again the exponent $0.6$ suggest the Golden mean $\phi $ as
possibly nearest value.

\section{Conclusions}

In this paper we have studied the effect of a stochastic
self-similar and fractal Universe on some physical quantities and
relations. By using the Virial Theorem and the proposed scale
invariant law, we derived the time dependence of $G$. We verify
the agreement between the theoretical value of the cosmological
constant coming from the presented stochastic self-similar
Universe and the observed one. Also the homogeneity of the
Universe appears in connection with the present law. Also using
the Virial Theorem, the evaluated $G(N,t)$ and the Hubble's law we
found an accelerating Universe similar to what has shown by the
recent observations on the SNe Ia. In addition, the exponent of
the correlation function can be explained in the context of the
stochastic self-similar Universe, equivalently as in El Naschie
$\varepsilon^{(\infty)}$ Cantorian space-time. Our model allows us
to realize an actual segregated Universe according to the
observations. Thanks to the relation $R=lN^{\alpha }$, we have a
link between the actual Universe, as observed, and its primordial
phase, when quantum and relativistic laws were in comparison with
gravity.

Relation (3') appears interesting not only because it allows us to
obtain the exact dimensions of self-gravitating systems, but it is
scale invariant. It is interesting to note that the observations
on the large--scale structures and the Random Walk relation
suggest $\alpha =3/2=1.5$ as best value (in agreement with El
Naschie's $E$-infinity Cantorian space-time, the Golden Mean and
the Fibonacci numbers). All these results confirm the fractality
of power law (1), which tends to be a more like a general theory.
In a certain sense, gravity was analyzed as a statistical property
of space-time and the random processes in it.

Thanks to these results we can conclude that  the fractal power
law represent a fractal Universe.

\subsubsection*{Acknowledgements}

The author wish to thank S.Capozziello for comments and
discussions.

\end{document}